\documentclass[twocolumn,showpacs,preprintnumbers,superscriptaddress,prl,amsmath,amssymb]{revtex4-1}

\usepackage[pdftex]{graphicx}
\usepackage{CJK}
\usepackage{dcolumn}% Align table columns on decimal point
\usepackage{bm}% bold math

\begin{document}
\begin{CJK*} {GB} {}
\CJKfamily{gbsn}
%\CJKfamily{min}
%\CJKfamily{mj}

\title{Anisotropic magnetoelastic coupling in single-crystalline CeFeAsO \\
as seen via high-resolution x-ray diffraction}

\author{H.-F. Li} %(À·å)}
\email{h.li@fz-juelich.de} \affiliation{Ames Laboratory, U.S. DOE, Ames, Iowa 50011, USA} \affiliation{J$\ddot{u}$lich Centre for Neutron Science JCNS,
Forschungszentrum J$\ddot{u}$lich GmbH, Outstation at Institut Laue-Langevin (ILL), B.P. 156, 38042 Grenoble Cedex 9, France}
\author{J.-Q. Yan}
\affiliation{Ames Laboratory, U.S. DOE, Ames, Iowa 50011, USA} \affiliation{Department of Materials Science and Engineering, University of Tennessee,
Knoxville, Tennessee 37996, USA} \affiliation{Materials Science and Technology Division, Oak Ridge National Laboratory, Oak Ridge, Tennessee 37831, USA}
\author{J. W. Kim}
\affiliation{Advanced Photon Source, Argonne National Laboratory, Argonne, Illinois 60439, USA}
\author{R. W. McCallum}
\affiliation{Ames Laboratory, U.S. DOE, Ames, Iowa 50011, USA} \affiliation{Department of Materials Science and Engineering, Iowa State University, Ames, Iowa 50011, USA}
\author{T. A. Lograsso}
\affiliation{Ames Laboratory, U.S. DOE, Ames, Iowa 50011, USA}
\author{D. Vaknin}
\affiliation{Ames Laboratory, U.S. DOE, Ames, Iowa 50011, USA} \affiliation{Department of Physics and Astronomy, Iowa State University, Ames, Iowa 50011,
USA}

\date{\today}

\begin{abstract}
Single-crystal synchrotron X-ray diffraction studies of CeFeAsO reveal strong anisotropy in the charge correlation lengths along or perpendicular to the
in-plane antiferromagnetic (AFM) wave-vector at low temperatures, indicating an anisotropic two-dimensional magnetoelastic coupling. The high-resolution
setup allows to distinctly monitor each of the twin domains by virtue of a finite misfit angle between them that follows the order parameter. In addition,
we find that the in-plane correlations, above the orthorhombic (O)-to-tetragonal (T) transition, are shorter than those in each of the domains in the AFM
phase, indicating a distribution of the in-plane lattice constants. This strongly suggests that the phase above the structural O-to-T transition is
virtually T with strong O-T fluctuations that are probably induced by spin fluctuations.
\end{abstract}


\maketitle
\end{CJK*}

\begin{figure} \centering \includegraphics [width = 0.47\textwidth] {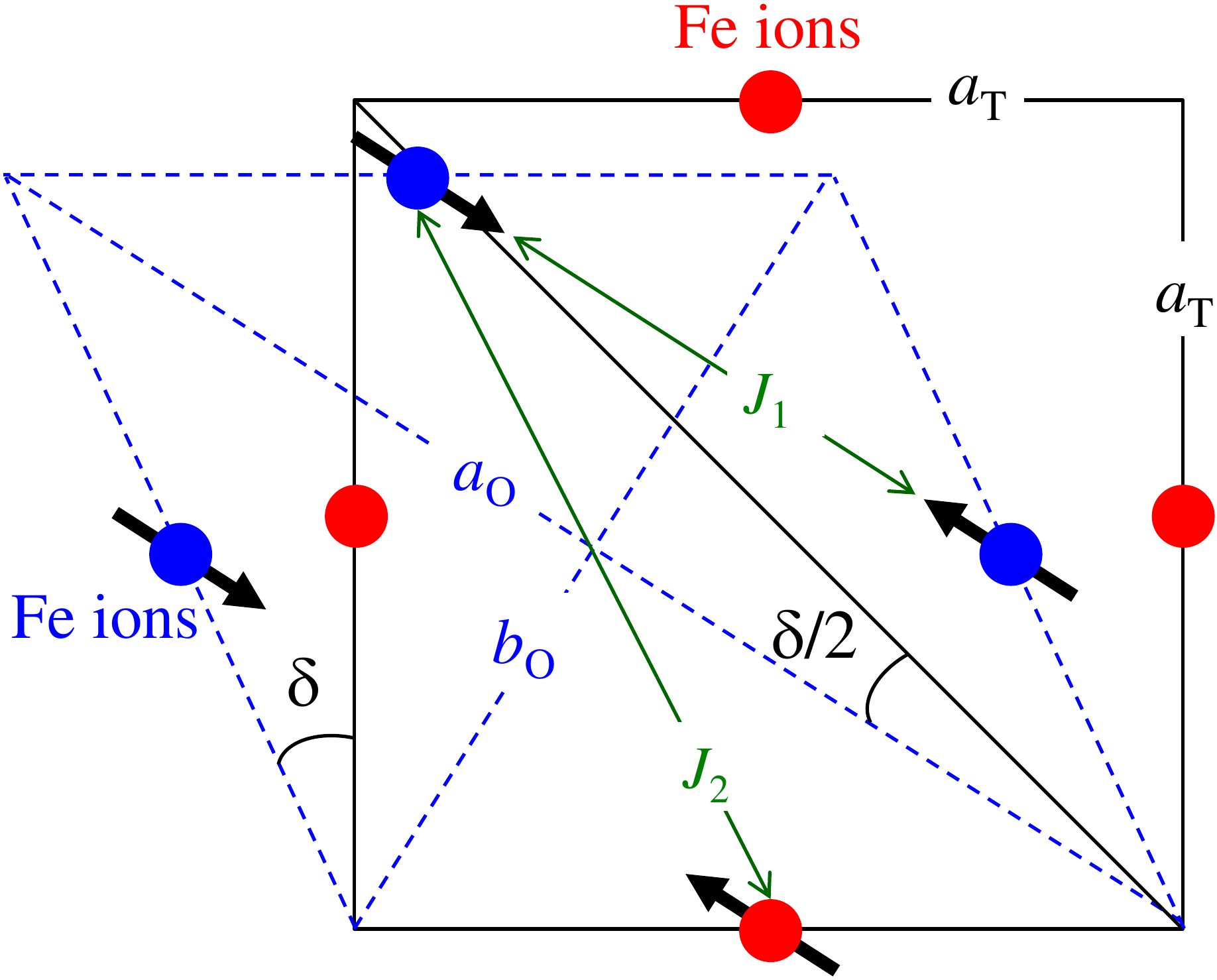}
\caption{(color online) Schematic illustration of the formation of twin domains when the T structure transfers into the O phase in CeFeAsO. The O distortion
in pnictides proceeds by shearing the T planar-square into two rhombuses with angles $\frac{\pi}{2} \pm \delta$ (preserving the length of
the square), creating twin domains. $\delta$ is the shearing angle. Here we just show one rhombus for clarity. $J_1$ and $J_2$ represent the NN and NNN AFM
exchange interactions, respectively. $a_\texttt{T}$ (T symmetry), $a_\texttt{O}$ and $b_\texttt{O}$ (O symmetry) are the in-plane lattice constants. The
arrows on the Fe ions represent the spins of iron ions.} \label{Figure1-1}
\end{figure}

Understanding the strong magnetoelastic coupling observed in the parent ferropnictides is pivotal to unraveling the mechanism that makes these systems
superconducting (SC) by chemical doping \cite{Kamihara2008,Chen2008,Ren2008}. This strong coupling is manifested in the emergence of a stripe-type
antiferromagnetic (AFM) phase that is stable in an orthorhombic (O) phase that results from shearing distortions of a high-temperature tetragonal (T) phase
as illustrated in Fig.\ \ref{Figure1-1}. The magnetic and structural transitions occur simultaneously in the $A$Fe$_2$As$_2$ ($A = $ Ca, Sr and Ba,
$\texttt{"}$122$\texttt{"}$) systems \cite{Yan2008, Goldman2008,Li2009-1}, while the magnetic ordering transition temperature ($T_\texttt{N}$) is in a
range of $\sim$ 6-18 K below the structural transition temperature ($T_\texttt{S}$) and strongly dependent on the sample quality in the $Ln$FeAsO ($Ln=$
lanthanide element, $\texttt{"}$1111$\texttt{"}$) family \cite{Jesche2010}. The spin-lattice coupling has also been implied in the interpretation of spin
dynamics as well as its influence on the structures of these systems. Inelastic neutron scattering studies from single-crystal CaFe$_2$As$_2$
\cite{Diallo2009, Zhao2009} and polycrystalline LaFeAsO \cite{Ishibashi2008} showed that spin fluctuations persist above $T_\texttt{N}$ up to at least room
temperature. It was argued that such fluctuations introduce dynamic disorder of the O/T phases, so that finite orthorhombicity and tetragonality may exist
above and below $T_\texttt{S}$, respectively \cite{Li2009-1, Jesche2010, Loudon2010}. Furthermore, the magnetic and structural transitions can be
simultaneously tuned by chemical substitutions, e.g., suppressing the AFM and O phases and setting in the SC state \cite{Paglione2010}.

\begin{figure} \centering \includegraphics [width = 0.47\textwidth] {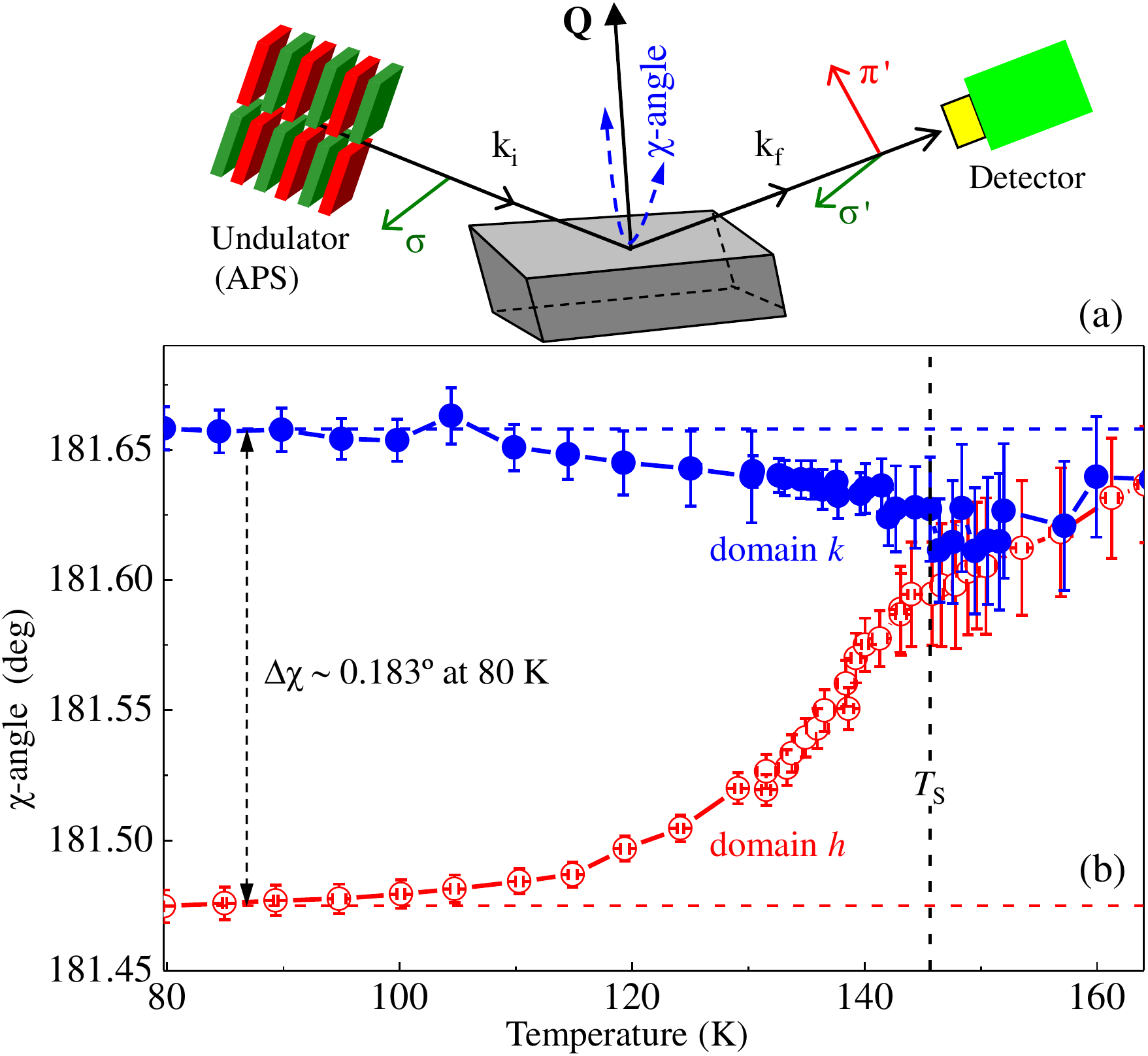}
\caption{(color online) (a) Illustration of the X-ray diffraction setup at the 6-ID-B beamline at APS at Argonne National Laboratory.
$\chi$-angle indicates the directions for tilting the CeFeAsO single crystal with respect to \textbf{Q}. (b) Temperature evolution of the $\chi$-angle
of the \emph{h} and \emph{k} domains. The vertical dashed line implies the structural O-T transition temperature \emph{T}$_\texttt{S}$. The
horizontal dashed lines indicate the $\chi$-angle difference between the \emph{h} and \emph{k} domains at 80 K. The solid lines are guides
to the eye.}
\label{Figure2-2}
\end{figure}

\begin{figure*} \centering \begin{minipage} [p] {0.64\linewidth} \includegraphics[width = 0.99\textwidth] {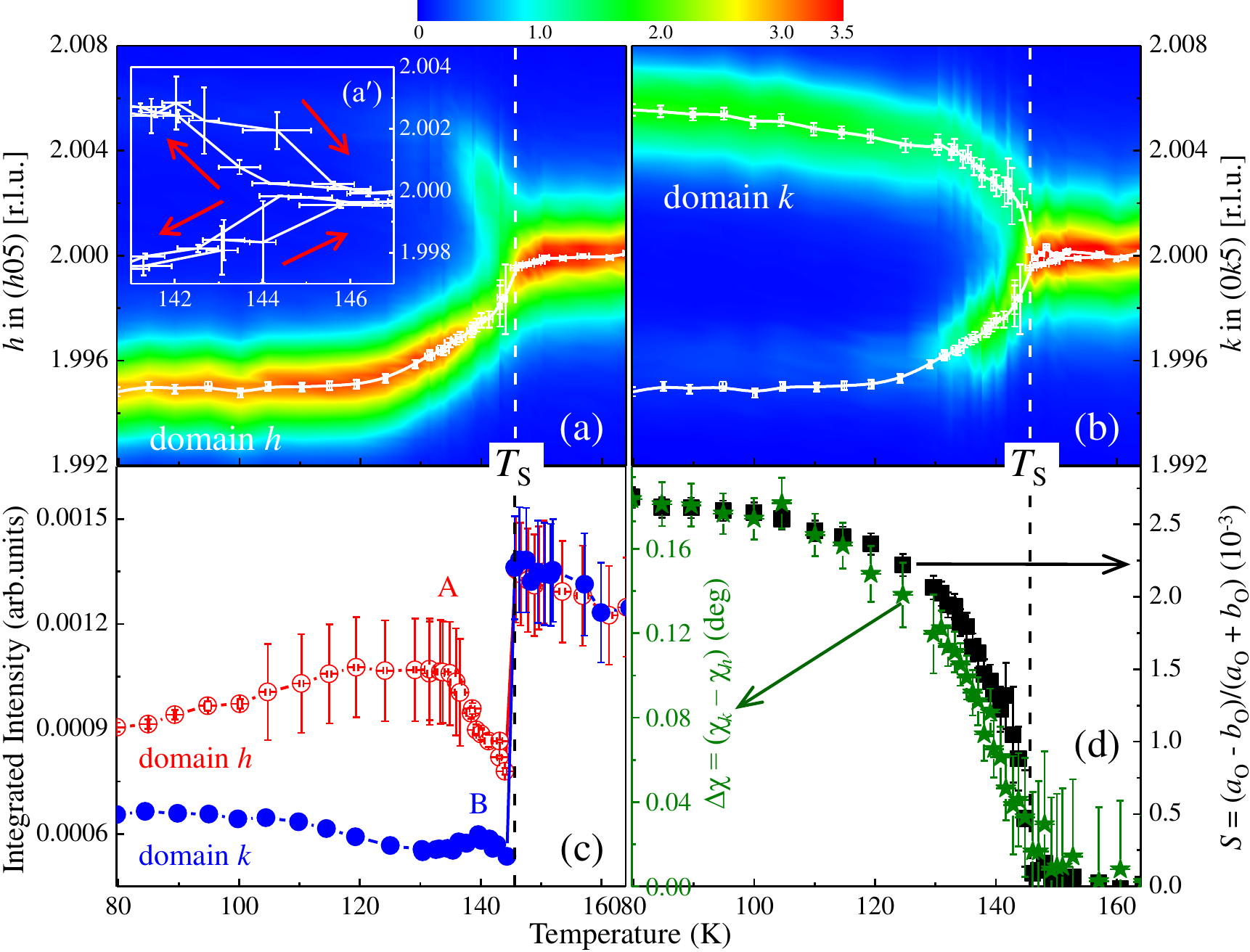}
\end{minipage}\hfill
\begin{minipage} [p] {0.34\linewidth}
\caption{(color online) Temperature evolution of (a) the \emph{h} domain and (b) the \emph{k} domain. The solid line in (a) and lower solid line in (b)
indicate the Lorentzian-fit-center of the longitudinal \emph{h} scans at \textbf{Q} = (\emph{h}05)$_\texttt{O}$. The upper solid line in (b) indicates the
Lorentzian-fit-center of the longitudinal \emph{k} scans at \textbf{Q} = (0\emph{k}5)$_\texttt{O}$. (c) Temperature evolutions of the integrated
intensities of the longitudinal \emph{h} and \emph{k} scans at \textbf{Q} = (\emph{h}05)$_\texttt{O}$ and (0\emph{k}5)$_\texttt{O}$, respectively, and (d)
the O distortion (squares) (right) in the crystalline $ab$ plane, namely, $S \equiv (a_\texttt{O} - b_\texttt{O})/(a_\texttt{\texttt{O}} + b_\texttt{O})$,
and the domain-angle misfit (stars) (left), namely, $\Delta\chi = \chi_k - \chi_h$. (a$'$) $\sim$ 2 K temperature hysteresis effect of the
Lorentzian-fit-centers indicative of a first-order O-T structural transition. Points A and B in (c) indicate the kink positions probably associated with
the AFM transition, as described in the test. The vertical dashed lines in (a)-(d) indicate the structural O-T transition temperature
\emph{T}$_\texttt{S}$. The solid lines in (a)-(c) and (a$'$) are guides to the eye.}
\end{minipage}
\label{Figure3-3}
\end{figure*}

The magnetic structure as schematically shown in Fig.\ \ref{Figure1-1} and spin dynamics indicate a strong magnetic frustration due to the competition
between the nearest-neighbor (NN) and the next-NN (NNN) AFM exchange couplings (referred to as the $J_1$-$J_2$ model) that produces a stripe-type AFM
structure for $J_2 > J_1$/2 in the parent pnictides. One of the consequences of the magnetic frustration is the possible emergence of nematic degrees of
freedom \cite{Chandra1990} that can give rise to a short-range O order above $T_\texttt{S}$ \cite{Fang2008, Xu2008, Fernandes2009}. However, directly
probing nematic fluctuations is nontrivial due to the difficulties in unequivocally decoupling their effects from that of the twin domains \cite{Chu2010}
as well as the fact that the magnetic field fluctuations associated with them average to zero. The roles of nematic fluctuations in shaping the magnetic
and structural transitions and in the pairing-mechanism that leads to superconductivity in the iron arsenides are under intense debate \cite{Fernandes2009,
chuang2010}.

Here we report on high-resolution synchrotron X-ray diffraction studies of a CeFeAsO single crystal that enable us to separately monitor the development of
each of the twin domains in this system. In particular, we examine the in-plane charge correlations as a function of temperature to gain insight on the
two-dimensional (2D) coupling between the structural and magnetic properties of this typical ferropnictide.

CeFeAsO single crystals were synthesized in NaAs flux at ambient pressure as described previously \cite{Yan2009}. Crystal quality was characterized by Laue
backscattering, X-ray powder-diffraction, heat-capacity, magnetization, and resistivity measurements. A large as-grown plate-like CeFeAsO single crystal
($\sim$ 5 mg) with the crystallographic \emph{c} axis perpendicular to its surface was selected. The X-ray diffraction studies were carried out on the
six-circle diffractometer of the 6-ID-B (XOR/MU) beamline at the Advanced Photon Source (APS) at Argonne National Laboratory. The X-ray energy throughout
the experiment was kept at \emph{E} = 7.1000(5) keV. The incident radiation was linearly $\sigma$ polarized with a spatial cross-section of 1.0 mm
(horizontal) $\times$ 0.25 mm (vertical). The scattering geometry is shown in Fig. 2(a), where the $\chi$-angle represents a relative tilting of the sample
with respect to \textbf{Q}. In this configuration, charge scattering does not change the polarization of the scattered photons and occurs in the
$\sigma$-$\sigma'$ scattering channel. The mosaic of the single crystal is $\sim$ 0.09$^\circ$ full width at half maximum (FWHM) as measured on the charge
(205)$_\texttt{O}$ Bragg reflection at 80 K. The sample was mounted at the end of the cold-finger of a displex cryogenic refrigerator with \emph{ac} plane
as the scattering plane and was measured at a temperature range of $\sim$ 80 to 170 K. We note that the (\emph{HKL})$_\texttt{T}$ indices for the T
symmetry correspond to the O reflection (\emph{hkl})$_\texttt{O}$ based on the relations of $h = H + K, k = H - K$, and $l = L$.
\begin{figure} \centering \includegraphics [width = 0.47\textwidth] {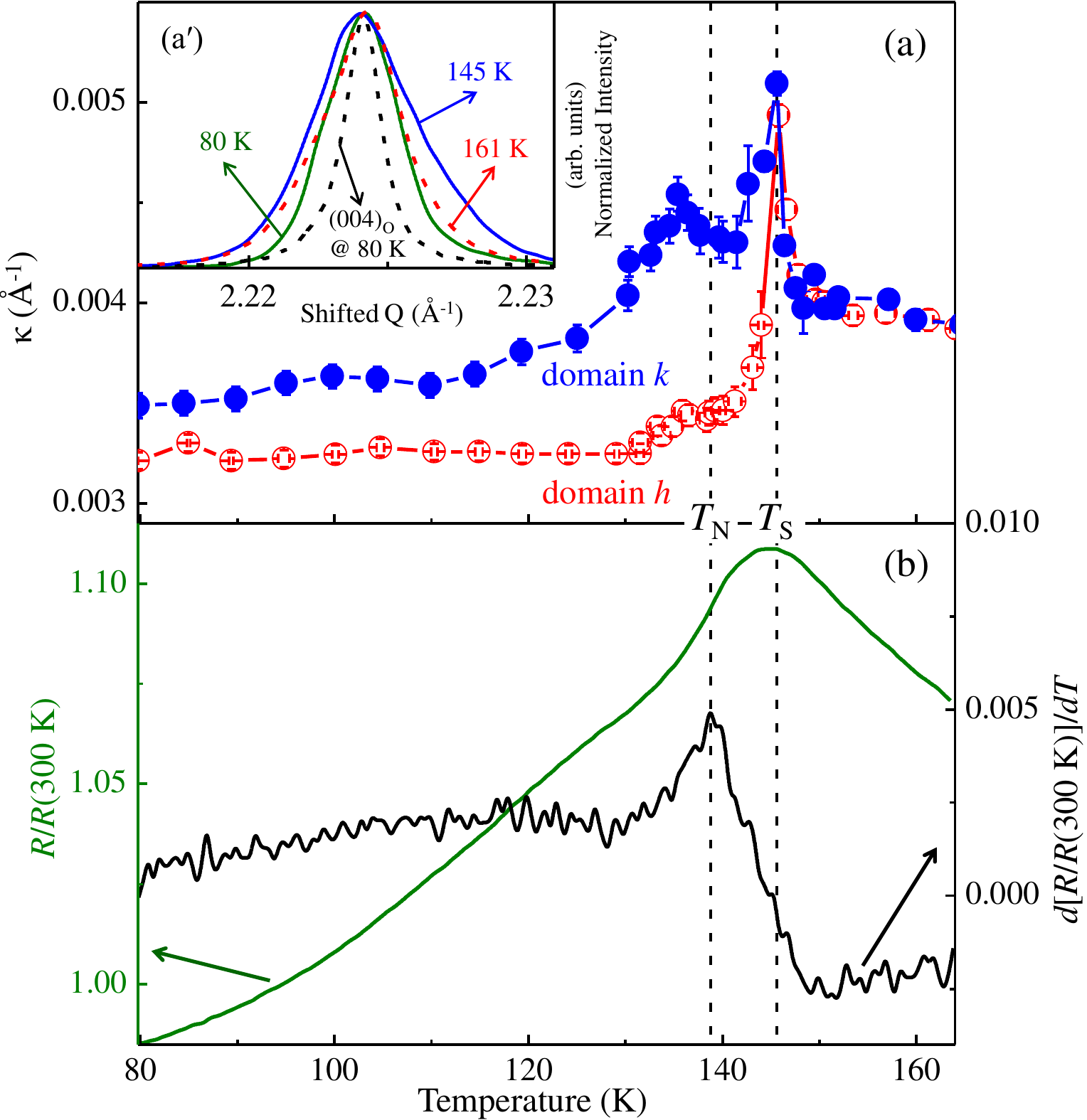}
\caption{(color online) Temperature evolutions of (a) the FWHM $\kappa$ of the longitudinal \emph{h} and \emph{k} scans at \textbf{Q} = (\emph{h}05)$_\texttt{O}$
and (0\emph{k}5)$_\texttt{O}$, respectively, and (b) the resistivity measurement (left) and its first derivative (right) indicate the temperatures of the AFM
transition of iron ions ($T^{\texttt{Fe}}_{\texttt{N}}$) as well as the O-T structural transition ($T_\texttt{S}$). (a$'$) shows the longitudinal \emph{h}
scans at \textbf{Q} = (\emph{h}05)$_\texttt{O}$ and at temperatures of 80, 145 and 161 K. To estimate the instrumental resolution effect, we also showed the
corresponding \emph{l}-scan of the (004)$_\texttt{O}$ reflection at 80 K. The Q values and observed intensities were shifted and normalized, respectively, for
comparison. The vertical dashed lines in (a) and (b) indicate the AFM and the O-T structural transition temperatures \emph{T}$_\texttt{N}$ and \emph{T}$_\texttt{S}$, respectively. The solid lines in (a) are guides to the eye.}
\label{Figure4-4}
\end{figure}

Upon cooling, at $T_\texttt{S}$, the T ($P4/nmm$) structure transfers into the O ($Cmma$) one. This is accompanied by an appearance of twin domains, e.g.,
the Bragg $(HK0)_\texttt{T}$ reflection in the T symmetry splits into twined $(H + K, H - K, 0)_\texttt{O}/(H - K, H + K, 0)_\texttt{O}$ ones in the O
symmetry. To obtain the angle misfit between the twin domains, we monitored the (205)$_\texttt{O}$/(025)$_\texttt{O}$ reflections. Figure 2(b) shows the
temperature dependence of the $\chi$-angle [as illustrated in Fig. 2(a)] of both reflections. We find that for the \emph{k} domain, represented by the
(025)$_\texttt{O}$ reflection, the $\chi$-angle has no appreciable change within statistics in the investigated temperature range of $\sim$ 80-164 K;
whereas for the \emph{h} domain, represented by the (205)$_\texttt{O}$ reflection, the $\chi$-angle gradually increases below $\sim$ 130 K upon warming,
and then steeply increases until merging into that of the \emph{k} domain at $T_\texttt{S}$. The measured $\chi$-angle difference, $\Delta\chi$ = $\chi_k -
\chi_h$, is $\sim$ 0.183$^\circ$ at 80 K as indicated in Fig. 2(b).

The separation of the twin domains below $T_\texttt{S}$ allows us to practically focus on an untwinned single crystal and follow the evolutions of the
lattice constants and the correlations along the \emph{a} and \emph{b} axes. Figures 3(a) and 3(b) show the temperature dependence of the (\emph{h}05) and
(0\emph{k}5) reflections from the \emph{h} and \emph{k} domains, respectively, indicating $T_\texttt{S} = 145(1) $ K. The integrated intensities of the
longitudinal \emph{h} and \emph{k} scans from both domains shown in Fig. 3(c) exhibit two features associated with $T_\texttt{S}$ and $T_\texttt{N}$. The
dramatic jump in the intensity at $T_\texttt{S}$ is clear evidence that the structural transition is first-order in character. This is also manifested by a
$\sim$ 2 K hysteresis effect as shown in Fig. 3(a$'$), consistent with similar observations in other pnictides \cite{Goldman2008,Li2009-1}. The other
feature in Fig. 3(c), labeled by A and B, is more subtle and may result from the magnetic transition as discussed below, providing evidence for the
coupling between lattice and spin orders.

Figure 3(d) shows the temperature dependence of the O strain $S$ and the misfit-angle difference $\Delta\chi = \chi_k - \chi_h$. The value of the strain
\emph{S} $\sim$ 2.7 $\times$ 10$^{-3}$ at 80 K in CeFeAsO [$S \equiv (a_\texttt{O}-b_\texttt{O})/(a_\texttt{O}+b_\texttt{O})$, where $a_\texttt{O}$ and
$b_\texttt{O}$ are the O lattice constants] is almost half the corresponding values in SrFe$_2$As$_2$ \cite{Li2009-1} and CaFe$_2$As$_2$ \cite{Goldman2008}
compounds. This indicates that the out-of-plane coupling in the $\texttt{"}$1111$\texttt{"}$ system is much weaker than that in the
$\texttt{"}$122$\texttt{"}$ family. The temperature dependence of the misfit-angle difference coincides well with that of the O strain, indicating a close
relationship between microscopic ($S$) and macroscopic ($\Delta\chi$) parameters. This demonstrates that $\Delta\chi$ can serve yet as another probe to
monitor the order parameter of the O-T structural transition.

The most remarkable observation in this study is the temperature variation in the peak-linewidth (i.e., the FWHM = $\kappa$) for both twin domains as shown
in Fig. 4(a). This linewidth obtained from the longitudinal \emph{h} and \emph{k} scans extends beyond the instrumental resolution as demonstrated by
comparing it with that of the corresponding \emph{l}-scans of the (004)$_\texttt{O}$ reflection [Fig. 4(a$'$)], and is thus inversely proportional to the
intrinsic in-plane charge correlation lengths. At low temperatures ($< T_\texttt{N}$), the charge correlation length along the $h$ direction is
significantly larger than that along the $k$ direction, indicating that the \emph{h}-domain is majority consistent with the stronger intensity observed for
the \emph{h}-domain in Fig. 3(c). This agrees well with the fact that the AFM interaction (inter-stripe) in this system is much stronger than the effective
FM one (intra-stripe), which results from the competing NNN interactions [$J_2$ as illustrated in Fig. 1] that introduce frustration in the magnetic
system. This charge-correlation-length anisotropy is also consistent with recent observations in an inelastic neutron scattering study of SC
Ba(Fe$_{0.926}$Co$_{0.074}$)$_2$As$_2$ that show a similar anisotropy in the spin correlation lengths \cite{Li2010}. The much smaller FWHM of the
(004)$_\texttt{O}$ reflection compared to that of the \emph{h}-scans of the (205)$_\texttt{O}$ reflection at 80 K [Fig. 4(a$'$)] in turn demonstrates that
the spin-lattice coupling is 2D, and thus the magnetic exchange along the \emph{l}-direction is very weak. With the increase in temperature, the anisotropy
becomes more and more pronounced, and two prominent peak-like features are observed at $\sim$ 135 and $\sim$ 145 K which we associate with the stripe-type
AFM and the O-T transitions, respectively. The relatively large broadening at $T_\texttt{S}$ is due to the genuine increase in linewidth with temperature
and the fact that the splitting may not be resolvable, and thus was treated as a single peak. Figure 4(b) shows the temperature dependence of the
normalized in-plane resistivity and its first derivative, clearly showing anomalies at $T_\texttt{S}$ and $T_\texttt{N}$, consistent with the observations
in Fig. 4(a).

In the T symmetry, above $T_\texttt{S}$, the linewidths are significantly larger than those at low temperatures. Similar observations have been reported in
powder-diffraction studies of LaFeAsO \cite{Li2010LaFeAsO, Qureshi2010}. This indicates that what is measured is in fact a distribution in the \emph{d}
spacings due to the fluctuating O/T structures, which may not be resolved on the time scale and precision of the instrument. This seems to be an averaged-T
phase with O/T fluctuations that are probably induced by the strong spin fluctuations typical in these systems \cite{Diallo2009}. It should be noted that
in general the correlation length diverges close to the O-T transition. However, the correlation lengths observed here are at their local minima close to
the transition, characteristic of a martensitic-like transition, due to the shearing distortion in these systems \cite{Loudon2010}.

The behavior of the asymmetric FWHM as shown in Fig. 4(a) may imply a remnant orthorhombicity above $T_\texttt{S}$. While the gradual decrease of the FWHM
of the \emph{k} domain below $\sim$ 130 K may indicate a remnant tetragonality below $T_\texttt{S}$. It is possible that the remnant tetragonality follows
the \emph{k} domain because both lattice constants $b_\texttt{O}$ and $b_\texttt{T}$ decrease upon cooling. Therefore, we argue that both T and O phases
may coexist dynamically in a certain temperature range around $T_\texttt{S}$, most likely due to the strong magnetic fluctuations, and perhaps, due to spin
nematic degrees of freedom \cite{Fang2008,Jesche2010,Li2010}. This scenario may explain the small temperature hysteresis effect [Fig. 3(a$'$)] and the
gradual increase of the order parameters in the first-order structural O-T transition below $T_\texttt{S}$ [Fig. 3(d)].

Our discussion above assumes that the variations in the linewidths are purely from correlations. Another alternative is the domain-size effect, in which
the variations indicate that in the critical fluctuation regime, magnetoelastic coupling already nucleates anisotropic domains, giving rise to elongated
domains upon freezing consistent with the observation in electron-microscopy studies \cite{Loudon2010}.

It is interesting to note that the structural O-T transition temperature $T_\texttt{S} = 145(1)$ K determined in Fig. 3(d) is also reflected in the
linewidths [Fig. 4(a)] and in the integrated intensities [Fig. 3(c)]. This shows that the temperature variation of integrated intensities could be useful
in determining the O-T structural transition temperature in these systems \cite{Li2009-1}. Similar observations by high-resolution X-ray diffraction were
reported for TbVO$_4$ and TbAsO$_4$ \cite{Rule2008}, where the T-to-O structural transition is driven by the cooperative Jahn-Teller distortion rather than
spin fluctuations.

To summarize, we demonstrate that with high-resolution synchrotron X-ray diffraction, the twin domains observed in single-crystal CeFeAsO can be distinct,
allowing the practical study of individual untwinned crystal. Most importantly, we find that the charge correlations show a significant anisotropy along
and perpendicular to the stripe-type 2D AFM wave-vector. This is consistent with the anisotropic 2D spin correlations indicative of an anisotropic 2D
magnetoelastic coupling, and implies that the AFM ordering is the driving force for the structural T-to-O transition in this system. In addition, we show
that the misfit-angle difference of the O twin domains (a macroscopic variable) follows well the O strain (or shearing) order parameter, indicating that it
can be used to monitor the order parameter of the O-T structural transition. Furthermore, based on the observed charge correlation lengths of the O twin
domains and the T domain below and above $T_\texttt{S}$, respectively, we argue that at high temperatures the system is a virtually T phase that exhibits
strong O/T structural fluctuations. This fluctuating phase is probably induced by the strong spin fluctuations, most likely by the spin-nematic phase.

Research at Ames Laboratory is supported by the U.S. Department of Energy, Office of Basic Energy Sciences, Division of Materials Sciences and Engineering
under Contract No. DE-AC02-07CH11358. Use of the Advanced Photon Source at Argonne National Laboratory was supported by the U.S. Department of Energy,
Office of Science, Office of Basic Energy Sciences, under Contract No. DE-AC02-06CH11357.

\end{document}